\documentclass[pra,aps,twocolumns,oneside]{revtex4}

\usepackage{amsmath,amssymb,latexsym}
\usepackage{graphicx}

\usepackage{bm}
\def\QQfnmark#1{\footnotemark}

\begin{document}

\title[Complexity of D-dimensional hydrogenic systems]{Complexity of D-dimensional hydrogenic systems in position and momentum spaces}

\author{S.\ L\'opez-Rosa}

\email{slopez@ugr.es}
\affiliation{Departamento de F\'{\i}sica At\'{o}mica, Molecular y Nuclear and Instituto Carlos I de F\'{\i}sica Te\'orica y
Computacional, Universidad de Granada, 18071-Granada, Spain}

\author{D. Manzano}

\email{manzano@ugr.es}
\affiliation{Departamento de F\'{\i}sica At\'{o}mica, Molecular y Nuclear and Instituto Carlos I de F\'{\i}sica Te\'orica y
Computacional, Universidad de Granada, 18071-Granada, Spain}
\email{manzano@ugr.es}

\author{J.\ S.\ Dehesa}

\email{dehesa@ugr.es}
\affiliation{Departamento de F\'{\i}sica At\'{o}mica, Molecular y Nuclear and Instituto Carlos I de F\'{\i}sica Te\'orica y
Computacional, Universidad de Granada, 18071-Granada, Spain}

\begin{abstract}

The internal disorder of a D-dimensional hydrogenic system, which is strongly associated to the non-uniformity of the quantum-mechanical density of its physical states, is investigated by means of the shape complexity in the two reciprocal spaces. This quantity, which is the product of the disequilibrium or averaging density and the Shannon entropic power, is mathematically expressed for both ground and excited stationary states in terms of certain entropic functionals of Laguerre and Gegenbauer (or ultraspherical) polynomials. We emphasize the ground and circular states, where the complexity is explicitly calculated and discussed by means of the quantum numbers and dimensionality. Finally, the position and momentum shape complexities are numerically discussed for various physical states and dimensionalities, and the dimensional and Rydberg energy limits as well as their associated uncertainty products are explicitly given. As a byproduct, it is shown that the shape complexity of the system in a stationary state does not depend on the strength of the Coulomb potential involved.\\\\
{\it Accepted in physica A}
\end{abstract}

\pacs{87.70.Cf, 03.65.-w, 87.19.-lo, 31.15.-p}

\maketitle
\newpage

\section{Introduction}

The hydrogenic system (i.e., a negatively-charged particle moving around a positively-charged core which electromagnetically binds it in its orbit) with dimensionality $D\geq 1$, plays a central role in D-dimensional quantum physics and chemistry \cite{H93, A00}. It includes not only a large variety of three-dimensional physical systems (e.g., hydrogenic atoms and ions, exotic atoms, antimatter atoms, Rydberg atoms,…) but also a number of nanoobjects so much useful in semiconductor nanostructures (e.g., quantum wells, wires and dots) \cite{H05, L07} and quantum computation (e.g., qubits) \cite{N00, D03}. Moreover it has a particular relevance for the dimensional scaling approach in atomic and molecular physics \cite{H93} as well as in quantum cosmology \cite{A05} and quantum field theory \cite{W80, I06}. Let us also say that the existence of hydrogenic systems with non standard dimensionalities has been shown for $D<3$ \cite{L07} and suggested for $D>3$ \cite{B99}. We should also highlight the use of D-dimensional hydrogenic wavefunctions as complete orthonormal sets for many-body problems \cite{AA, ACC} in both position and momentum spaces, explicitly for three-body Coulomb systems (e.g. the hydrogen molecular ion and the helium atom); generalizations are indeed possible in momentum-space orbitals as well as in their role as Sturmians in configuration spaces.

The internal disorder of this system, which is manifest in the non-uniformity quantum-mechanical density and in the so distinctive hierarchy of its physical states, is being increasingly investigated beyond the root-mean-square or standard deviation (also called Heisenberg measure) by various information-theoretic elements; first, by means of the Shannon entropy \cite{Y94, D98, D01} and then, by other individual information and/or spreading measures as the Fisher information and the power and logarithmic moments \cite{V00}, as it is described in Ref. \cite{D08} where the information theory of D-dimensional hydrogenic systems is reviewed in detail. Just recently, further complementary insights have been shown to be obtained in the three-dimensional hydrogen atom by means of composite information-theoretic measures, such as the Fisher-Shannon and the shape complexity \cite{S08, L08}. In particular, Sa\~nudo and Lopez-Ruiz \cite{S08} have found some numerical evidence that, contrary to the energy, both the Fisher-Shannon measure and the shape complexity in the position space do not present any accidental degeneracy (i.e. they do depend on the orbital quantum number $l$); moreover, they take on their minimal values at the circular states (i.e., those with the highest $l$). In fact, the position Shannon entropy by itself has also these two characteristics as it has been numerically pointed out long ago \cite{Y94}, where the dependence on the magnetic quantum number is additionally studied for various physical states.

The shape complexity \cite{C02} occupies a very special position not only among  the composite information-theoretic measures in general, but also within the class of measures of complexity. This is because of the following properties: (i) invariance under replication, translation and rescaling transformations, (ii) minimal value for the simplest probability densities (namely, e.g. uniform and Dirac's delta in one-dimensional case), and (iii) simple mathematical structure: it is given as the product of the disequilibrium or averaging density and the Shannon entropy power of the system.

In this work we provide the analytical methodology to calculate the shape complexity of the stationary states of the  D-dimensional hydrogenic system in the two reciprocal position and momentum spaces and later we apply it to a special class of physical states which includes the ground state and the circular states (i.e. states with the highest hyperangular momenta allowed within a given electronic manifold). First, in Section 2, we briefly describe the known expressions of the quantum-mechanical density of the system in both spaces. In Section 3 we show that the computation of the two shape complexities for arbitrary D-dimensional hydrogenic stationary states boils down to the evaluation of some entropic functionals of Laguerre and Gegenbauer polynomials. To have the final expressions of these complexity measures in terms of the dimensionality D and the quantum numbers characterizing the physical state under consideration, we need to compute the values of these polynomial entropic functionals what is, in general, a formidable open task. However, in Section 4, we succeed to do it for the important cases of ground and circular states. It seems that for the latter ones the shape complexity has the minimal values, at least in the three-dimensional case as indicated above. It is also shown that our results always fulfil the uncertainty relation satisfied by the position and momentum shape complexities \cite{L08}. In Section V, the shape complexities are numerically studied and their dimensionality dependence is discussed. Finally, some conclusions are given.

\section{The D-dimensional hydrogenic quantum-mechanical densities}

Let us consider an electron moving in the D-dimensional Coulomb ($D \geqslant 2$) potential $V(\vec{r})=-\frac{Z}{r}$, where $\vec{r}=(r, \theta_1,\theta_2,...,\theta_{D-1})$ denotes the electronic vector position in polar coodinates. The stationary states of this hydrogenic system are described by the wavefunctions
\begin{equation}\nonumber
\Psi_{n,l,\left\lbrace \mu \right\rbrace }\left(\vec{r},t \right)=\psi_{n,l, \left\lbrace \mu \right\rbrace }\left(\vec{r}\right) \exp \left(- i E_n t\right),
\end{equation}
where $\left(E_n,\Psi_{n,l, \left\lbrace \mu \right\rbrace }  \right)$ denote the physical solutions of the Schr\"odinger equation of the system \cite{H93, A00, D08}. Atomic units $\left(\hslash=e=m_e=1 \right)$ are used throughout the paper. The energies are given by
\begin{equation} \label{energia}
E= -\frac{Z^2}{2\eta^2},\hspace{0.5cm} \text{ with } \hspace{0.5cm} \eta=n+\frac{D-3}{2}; \hspace{5mm} n=1,2,3,...,
\end{equation}
and the eigenfunctions can be expressed as
\begin{equation}\label{FunOndaR}
\Psi_{n,l, \left\lbrace \mu \right\rbrace }(\vec{r})=R_{n,l}(r) {\cal{Y}}_{l,\{\mu\}}(\Omega_{D-1}),
\end{equation}
where $(l,\left\lbrace \mu \right\rbrace)\equiv(l\equiv\mu_1,\mu_2,...,\mu_{D-1})\equiv(l,\{\mu\})$ denote the hyperquantum numbers associated to the angular variables $\Omega_{D-1}\equiv (\theta_1, \theta_2,...,\theta_{D-1} \equiv \varphi)$, which may have all values consistent with the inequalities $l\equiv\mu_1\geq\mu_2\geq...\geq \left|\mu_{D-1}\right|\equiv \left|m\right|\geq 0$. The radial function is given by
\begin{equation}\label{radial}
R_{n,l}(r)= \left( \frac{\lambda^{-D}}{2 \eta}\right)^{1/2}   \left[\frac{\omega_{2L+1}(\hat{r})}{\hat{r}^{D-2}}\right]^{1/2}{\tilde{\cal{L}}}_{\eta-L-1}^{2L+1}(\hat{r}),
\end{equation}
where ${\tilde{\cal{L}}}_{k}^{\alpha}(x)$ denotes the Laguerre polynomials of degree $k$ and parameter $\alpha$, orthonormal with respect to the weight function $\omega_\alpha (x)= x^\alpha e^{-x}$, and the grand orbital angular momentum hyperquantum number $L$ and the adimensional parameter $\hat{r}$ are
\begin{equation} \label{nL}
L=l+\frac{D-3}{2}, \hspace{0.5cm} l=0,1,2,... \hspace{0.5cm} \text{ and } \hspace{0.5cm} \hat{r}=\frac{r}{\lambda},\hspace{0.5cm} \text{ with } \hspace{0.5cm}\lambda=\frac{\eta}{2Z}.
\end{equation}

The angular part ${\cal{Y}}_{l,\{\mu\}}(\Omega_{D-1})$ is given by the hyperspherical harmonics \cite{A00,A93}
\begin{equation}\label{hyperesf}
{\cal{Y}}_{l,\{\mu\}}(\Omega_{D-1})=\frac{1}{\sqrt{2 \pi}} e^{im\varphi} \prod^{D-2}_{j=1} {\tilde{C}}_{\mu_{j}-\mu_{j+1}}^{\alpha_j+\mu_{j+1}}(\cos \theta_j) \left( \sin \theta_j\right)^{\mu_{j+1}},
\end{equation}
with $\alpha_j= \frac{1}{2} (D-j-1)$ and ${\tilde{C}}^{\lambda}_{k}(x)$ denotes the orthonormal Gegenbauer polynomials of degree $k$ and parameter $\lambda$.

Then, the quantum-mechanical probability density of the system in position space is
\begin{equation}\label{probdif}
\rho_{n,l,\left\lbrace \mu \right\rbrace} (\vec{r}) = \left|\Psi_{n,l,\left\lbrace \mu \right\rbrace} \left( \vec{r} \right)  \right|^2 = R_{n,l}^{2} (r) \left|{\cal{Y}}_{l,\left\lbrace \mu \right\rbrace} \left( \Omega_{D-1}\right)  \right|^2 ,
\end{equation}

In momentum space the eigenfunction of the system is \cite{A00, D08, DL07, AQ}
\begin{equation}\label{FunOndaP}
\tilde{\Psi}_{nl\{\mu\}}(\vec{p})={\cal{M}}_{n,l}(p){\cal{Y}}_{l\{\mu\}}(\Omega_{D-1}),
\end{equation}
where the radial part is
\begin{align}\label{momento}
{\cal{M}}_{n.l}(p)&= 2^{L+2} \left(  \frac{\eta} {Z}\right) ^{D/2}  \frac{(\eta \tilde{p})^l}{(1+\eta^2 {\tilde{p}}^2)^{L+2}} {\tilde{C}}^{L+1}_{\eta-L-1}\left(\frac{1-\eta^2 \tilde{p}^2}{1+\eta^2 \tilde{p}^2}\right)\\\nonumber
&=\left(  \frac{\eta} {Z}\right) ^{D/2} (1+y)^{3/2} \left(\frac{1+y}{1-y} \right)^{\frac{D-2}{4}} {\omega^{*}}^{1/2}_{L+1} (y) {\tilde{C}}^{L+1}_{\eta-L-1}(y),
\end{align}
with $y=\frac{1-\eta^2 \tilde{p}^2 }{1+\eta^2 \tilde{p}^2}$ and $\tilde{p}=\frac{p}{Z}$ (here the electron momentum $p$ is assumed to be expressed in units of $p_\mu$, where $p_{\mu_r}=\frac{\mu_r}{m_e}p_0=\mu_r$ m.a.u, since $m_e=1$ and the momentum atomic unit is $p_0=\frac{\hbar}{a_0}=\frac{m_e e^2}{\hbar}$; $\mu_r$ is the reduced mass of the system). The symbol ${\tilde{C}}^\alpha_m(x)$ denotes the Gegenbauer polynomial of order $k$ and parameter $\alpha$ orthonormal with respect to the weight funciont $\omega^{*}_{\alpha}=(1-x^2)^{\alpha-\frac{1}{2}}$ on the interval $\left[-1,+1 \right]$. The angular part is again an hyperspherical harmonic as in the position case, but with the angular variables of the vector $\vec{p}$. Then, one has the following expression
\begin{equation}\label{denmom}
\gamma(\vec{p})=\left|\tilde{\Psi}_{n,l,\{\mu\}}\left( \vec{p} \right) \right|^2={\cal{M}}^2_{n,l}(p) \left[ {\cal{Y}}_{l\{\mu\}}(\Omega_{D-1})\right]^2
\end{equation}
for the quantum-mechanical probability density of the system in momentum space.

\section{The shape complexity of the D-dimensional hydrogenic system}

Here we describe the methodology to compute the position and momentum shape complexity of our system in an arbitrary physical state characterized by the hyperquantum numbers $\left( \eta, \mu_1, ..., \mu_{D-1} \right)$. We show that the calculation of the position and momentum hydrogenic shape complexities ultimately reduce to the evaluation of some entropic functionals of Laguerre and Gegenbauer polynomials.

\subsection*{Position space}

The shape complexity $C \left[\rho \right]$ of the position probability density $\rho\left(\vec{r}\right)$ is defined \cite{C02} as
\begin{equation}\label{defCpos}
C\left[ \rho \right]= \left\langle   \rho \right\rangle \exp \left( S\left[ \rho \right] \right),
\end{equation}
where
\begin{equation}\label{Drho}
\left\langle   \rho \right\rangle = \int \rho^2 \left(\vec{r} \right) d \vec{r}
\end{equation}
and
\begin{equation}\label{Srho}
S\left[ \rho \right]= - \int \rho \left(\vec{r} \right)  \ln  \rho \left(\vec{r} \right)  d\vec{r}
\end{equation}
denote the first-order frequency moment (also called averaging density or disequilibrium, among other names) and the Shannon entropy of $\rho \left( \vec{r} \right)$, respectively. Then, this composite information-theoretic quantity measures the complexity of the system by means of a combined balance of the average height (as given by $ \left\langle \rho \right\rangle)$ and the total bulk extent (as given by $S \left[ \rho \right]$) of the corresponding quantum-mechanical probability density $\rho \left( \vec{r} \right)$.

Let us first calculate $\left\langle   \rho \right\rangle$. From (\ref{FunOndaR}) and (\ref{Drho}) one obtains that
\begin{equation}\label{Drho1}
\left\langle   \rho \right\rangle = \frac{2^{D-2}}{\eta^{D+2}} Z^D K_1 \left( D,\eta,L \right) K_2 \left( l, \left\lbrace \mu \right\rbrace \right),
\end{equation}
where
\begin{equation}\label{K1}
K_1 \left( D,\eta,L \right) = \int_{0}^{\infty} x^{-D-5} \left\lbrace \omega_{2L+1} (x) \left[ {\tilde{L}}^{2L+1}_{\eta-L-1} (x) \right]^2  \right\rbrace^2 dx
\end{equation}
and
\begin{equation}\label{K2}
K_2 \left( l, \left\lbrace \mu \right\rbrace \right) = \int_{\Omega} \left| {\cal{Y}}_{l\{\mu\}}\left(\Omega_{D-1}\right) \right|^4 d\Omega_{D-1}
\end{equation}

The Shannon entropy of $\rho \left( \vec{r} \right)$ has been recently shown \cite{DL08} to have the following expression
\begin{equation}\label{Srho1}
S \left[ \rho \right]= S \left[ R_{nl} \right] + S \left[ {\cal{Y}}_{l\{\mu\}} \right],
\end{equation}
with the radial part
\begin{align}\nonumber
S\left[ R_{n,l} \right]& = - \int_{0}^{\infty} r^{D-1} R_{n,l}^{2} (r) \log R_{n,l}^{2} dr \\\label{SR}
& = A(n,l,D) + \frac{1}{2 \eta} E_1\left[ \tilde{{\cal{L}}}^{2L+1}_{\eta-L-1} \right]-D \ln Z
\end{align}
and the angular part
\begin{align}\nonumber
S \left[{\cal{Y}}_{l,\left\lbrace \mu \right\rbrace} \right]& =- \int_{S_{D-1}} \left|{\cal{Y}}_{l,\left\lbrace \mu \right\rbrace} \left( \Omega_{D-1}\right)  \right|^2 \ln \left| {\cal{Y}}_{l,\left\lbrace \mu \right\rbrace} \left( \Omega^{'}_{D-1}\right)  \right|^2 d\Omega_{D-1} \\\label{SY}
& = B (l,\left\lbrace \mu \right\rbrace,D)+\sum^{D-2}_{j=1} E_2 \left[ \tilde{C}^{\alpha_j+\mu_{j+1}}_{\mu_j-\mu_{j+1}} \right],
\end{align}
where $A(n,l,D)$ and $B(l,\left\lbrace \mu \right\rbrace , D)$ have the following values 
\begin{equation}\nonumber
A(n,l,D)= -2l \left[\frac{2\eta-2L-1}{2 \eta} +\psi (\eta+L+1) \right]+\frac{ 3 \eta^2-L(L+1)}{\eta}+ -\ln \left[ \frac{2^{D-1}}{\eta^{D+1}} \right],
\end{equation}
and
\begin{align*}\nonumber
B (l,\left\lbrace \mu \right\rbrace,D)&= \ln 2\pi -2 \sum^{D-2}_{j=1} \mu_{j+1} \\ &\quad \times \left[\psi(2\alpha_j+\mu_j+\mu_{j+1})-\psi(\alpha_j+\mu_j)-\ln 2- \frac{1}{2 (\alpha_j+\mu_j)}\right],
\end{align*}
with $\psi(x) = \frac{\Gamma^{'}(x)}{\Gamma(x)}$ is the digamma function. The entropic functionals $E_i \left[ {\tilde{y}}_{n} \right]$, $i=1$ and $2$, of the polynomials $\left\lbrace  {\tilde{y}}_{n} \right\rbrace $, orthonormal with respect to the weight function $\omega(x)$, are defined \cite{D98, ADY} by
\begin{equation}\label{E1}
E_1 \left[{\tilde{y}}_n \right]=-\int_{0}^{\infty} x \omega(x) {\tilde{y}}^{2}_{n} (x) \ln {\tilde{y}}^{2}_{n}(x) dx,
\end{equation}
and
\begin{equation}\label{E2}
E_2 \left[{\tilde{y}}_n\right]=-\int_{-1}^{+1} \omega(x) {\tilde{y}}^{2}_{n} (x) \ln {\tilde{y}}^{2}_{n}(x) dx,
\end{equation}
respectively.

Finally, from Eqs. (\ref{defCpos}), (\ref{Drho1}) and (\ref{Srho1})-(\ref{SY}), we obtain the following value for the position shape complexity of our system:
\begin{align}\label{CposFinal}
C \left[ \rho \right] &= \frac{2^{D-2}}{\eta^{D+2}} K_1 \left(D, \eta, L \right) K_2 \left( L, \left\lbrace \mu \right\rbrace \right)\\ \nonumber
& \quad \times \exp \left[ A(n,l,D) + \frac{1}{2 \eta} E_1 \left[{\tilde{L}}_{\eta-L-1}^{2 L+1} \right] +S\left[ {\cal{Y}}_{l,\left\lbrace \mu \right\rbrace} \right]\right],
\end{align}
where the entropy of the hyperspherical harmonics $S\left[ {\cal{Y}}_{l,\left\lbrace \mu \right\rbrace} \right]$, given by Eq. (\ref{SY}), is controlled by the entropy of Gegenbauer polynomials $E_2 \left[{\tilde{C}}_k^{\alpha}\right]$ defined by Eq. (\ref{E2}). It is important to remark that the position complexity $C \left[\rho \right]$ does not depend on  the strength of the Coulomb potential, that is, on the nuclear charge $Z$.

\subsection*{Momentum space}

The shape complexity $C \left[ \gamma \right]$ of the momentum probability density $\gamma \left( \vec{p} \right)$ is given by
\begin{equation}\label{defCmom}
C\left[ \gamma \right]= \left\langle   \gamma \right\rangle \exp \left( S\left[ \gamma \right] \right),
\end{equation}
where the momentum averaging density $\left\langle \gamma \right\rangle$ can be obtained from Eq. (\ref{denmom}) as follows:

\begin{equation}\label{Cmom}
\left\langle   \gamma \right\rangle = \int \gamma^2 \left(\vec{p} \right) d \vec{p}= \frac{2^{4L+8} \eta^{D} }{Z^D} K_3 \left( D,\eta,L \right) K_2 \left( l, \left\lbrace \mu \right\rbrace \right),
\end{equation}
with $K_2$ is given by Eq. (\ref{K2}), and $K_3$ can be expressed as
\begin{equation}\label{K3}
K_3 \left( D,\eta,L \right) = \int_{0}^{\infty} \frac{y^{4l+D-1}}{(1+y^2)^{4L+8}} \left[ {\tilde{C}}^{L+1}_{\eta-L-1} (\frac{1-y^2}{1+y^2}) \right]^4  dy
\end{equation}

On the other hand, the momentum Shannon entropy $S\left[ \gamma \right]$ can be calculated in a similar way as in the position case. We have obtained that
\begin{align}\nonumber
S \left[ \gamma \right] & = - \int \gamma \left( \vec{p} \right) \ln \gamma \left( \vec{p} \right) d\vec{p} = S \left[ {\cal{M}}_{nl} \right] + S \left[ {\cal{Y}}_{l,\{\mu\}} \right] \\ \label{Sgamma}
&= F \left(n,l,D \right)+ E_2 \left[{\tilde{C}}_{\eta-L-1}^{L+1} \right]+D \ln Z+S\left[ {\cal{Y}}_{l,\left\lbrace \mu \right\rbrace} \right],
\end{align}
where $F\left(n,l,D \right)$ has been found to have the value
\begin{align}\label{F1}
F\left(n,l,D\right)&=-\ln \frac{\eta^D}{2^{2L+4}}-(2L+4) \left[ \psi(\eta+L+1)-\psi (\eta) \right]\nonumber \\
& \quad +\frac{L+2}{\eta}-(D+1) \left[1-\frac{2 \eta (2L+1)}{4\eta^2-1} \right]
\end{align}

Then, from Eqs. (\ref{defCmom}), (\ref{Cmom}) and (\ref{Sgamma}) we finally have the following value for the momentum shape complexity
\begin{align}\label{CmomFinal}
C \left[ \gamma \right] &= 2^{4 L+8} \eta^D  K_3 \left(D, \eta, L \right) K_2 \left( L, \left\lbrace \mu \right\rbrace \right)\\ \nonumber
& \quad \times \exp \left\lbrace  F(n,l,D) + E_2 \left[{\tilde{C}}_{\eta-L-1}^{L+1} \right] +S\left[ {\cal{Y}}_{l,\left\lbrace \mu \right\rbrace} \right] \right\rbrace 
\end{align}

Notice that, here again, this momentum quantity does not not depend on the nuclear charge Z. Moreover the momentum complexity $C \left[\rho \right]$ is essentially controlled by the entropy of the Gegenbauer polynomials $E_2 \left[ {\tilde{C}}_{k}^{\alpha} \right]$, since the entropy of hyperspherical harmonics $S\left[ {\cal{Y}}_{l,\left\lbrace \mu \right\rbrace} \right]$ reduces to that of these polynomials according to Eq. (\ref{SY}).

\section{Shape complexities of ground and circular states}

Here we apply the general expressions (\ref{CposFinal}) and (\ref{CmomFinal}) found for the position and momentum shapes complexities of an arbitrary physical state of the D-dimensional hydrogenic system, respectively, to the ground state $\left( n=1,\mu_i=0, \forall i=1...D-1 \right)$ and to the circular states. A circular state is a single-electron state with the highest hyperangular momenta allowed within a given electronic manifold, i.e. a state with hyperangular momentum quantum numbers $\mu_i=n-1$ for all $i=1,...,D-1$.

\subsection*{Ground state}

In this case $\eta-L-1=0$, so that the Laguerre polynomial involved in the radial wavefunction is a constant. Then, the probability density of the ground state in position space given by Eqs. (\ref{radial}), (\ref{hyperesf}) and (\ref{probdif}) reduces as follows:
\begin{equation}\label{denPosGS}
\rho_{g.s.}(\vec{r})= \left(\frac{2 Z}{D-1} \right)^D \frac{1 }{\pi^{\frac{D-1}{2}} \Gamma\left( \frac{D+1}{2} \right) } e^{-\frac{4Z}{D-1}r},
\end{equation}
which has been also found by various authors (see e.g. \cite{H93, A00}).

The expressions (\ref{Drho1})-(\ref{K2}), which provide the averaging density of arbitrary quantum-mechanical state, reduce to the value
\begin{equation}\label{DposGS}
\left\langle \rho_{g.s.} \right\rangle = \frac{Z^D}{(D-1)^D} \frac{1}{\pi^{\frac{D-1}{2}} \Gamma \left( \frac{D+1}{2} \right)}
\end{equation}
for the ground-state averaging density. Moreover, the angular part of the entropy is
\begin{equation}\label{SYgs}
S \left[{\cal{Y}}_{0,\left\lbrace 0 \right\rbrace  } \right]= \ln \frac{2 \pi^{D/2}}{\Gamma \left( \frac{D}{2} \right)},
\end{equation}
so that it is equal to  $\ln 2 \pi$ and $\ln 4 \pi$ for $D=2$ and $3$, respectively. Then, the formulas (\ref{Srho1})-(\ref{E2}) of the Shannon entropy of arbitrary physical state of our system simplify as
\begin{equation}\label{SposGS}
S \left[ \rho_{g.s.} \right]=\ln \left( \frac{(D-1)^D}{2^D} \pi^{\frac{D-1}{2}} \Gamma \left( \frac{D+1}{2} \right)  \right)+D-D \ln Z 
\end{equation}
for the ground-state Shannon entropy. Finally, from Eq. (\ref{CposFinal}) or from its own definition together with (\ref{DposGS})-(\ref{SposGS}) we obtain that the position shape complexity of D-dimensional hydrogenic ground state has the value
\begin{equation}\label{CposGS}
C \left[\rho_{g.s.} \right]= \left( \frac{e}{2}\right)^D
\end{equation}

In momentum space we can operate in a similar manner. First we have seen that the ground-state probability density is
\begin{equation}\label{denMomGS}
\gamma_{g.s.}(\vec{p})= \frac{(D-1)^D \Gamma \left(\frac{D+1}{2} \right) }{Z^D \pi^{\frac{D+1}{2}}} \frac{1}{\left( 1+\frac{(D-1)^2}{4} {\tilde{p}}^2 \right)^{D+1}},
\end{equation}
which has been also given by Aquilanti et al \cite{AQ}, among others. Then, we have found the values
\begin{equation}\label{DmomGS}
\left\langle \gamma_{g.s.} \right\rangle = \left( \frac{2D-2}{Z} \right)^D \frac{1}{\pi^{\frac{D+2}{2}}} \frac{\Gamma^2 \left( \frac{D+1}{2} \right) \Gamma \left(2+\frac{3D}{2} \right)}{\Gamma \left(2D+2 \right)}
\end{equation}
for the momentum averaging density, and
\begin{equation}\label{SmomGS}
S \left[ \gamma_{g.s.} \right]=\ln \frac{\pi^{\frac{D+1}{2}}}{(D-1)^{D} \Gamma\left(\frac{D+1}{2} \right) } + (D+1) \left[ \psi (D+1)-\psi \left( \frac{D}{2}+1 \right)  \right]+D \ln Z
\end{equation}
for the momentum Shannon entropy, directly from Eq. (\ref{DposGS}) or from Eqs. (\ref{Cmom})-(\ref{K3}) and (\ref{Sgamma})-(\ref{F1}), respectively. Finally. from Eq. (\ref{CmomFinal}) or by means of Eqs. (\ref{DmomGS})-(\ref{SmomGS}) we have the following value
\begin{equation}\label{CmomGS}
C \left[\gamma_{g.s.} \right]= \frac{2^D \Gamma \left( \frac{D+1}{2}\right) \Gamma \left(2+\frac{3 D}{2} \right) } {\pi^{1/2} \Gamma \left( 2D+2\right)} \exp \left\lbrace  (D+1) \left[ \psi \left( D+1 \right)-\psi \left( \frac{D+2}{2} \right)  \right]  \right\rbrace 
\end{equation}
for the ground-state D-dimensional hydrogenic shape complexity in momentum space. In particular, this quantity has the values
\begin{equation}\nonumber
C_2 (\gamma_{g.s.})= \frac{2 e^{3/2}}{5}=1.7926
\end{equation}
\begin{equation}\nonumber
C_3 (\gamma_{g.s.})=\frac{66}{e^{10/3}}=2.3545
\end{equation}
\begin{equation}\nonumber
C_4 (\gamma_{g.s.})=\frac{e^{35/12}}{6}=3.0799
\end{equation}
for the hydrogenic system with dimensionalities $D=2,3$ and $4$, respectively. Let us here mention that the three-dimensional value agrees with that calculated in \cite{C02}.

\subsection*{Circular states}

Following a parallel process with circular states, we have obtained
\begin{equation}\nonumber
\rho_{c.s.}(\vec{r}) =\frac{2^{D+2-2n} Z^D  }{\pi^{\frac{D-1}{2}}(2n+D-3)^{D} \Gamma(n) \Gamma \left(n+\frac{D-1}{2} \right)  } e^{-\frac{r}{\lambda} } \left( \frac{r}{\lambda} \right)^{2n-2} \prod^{D-2}_{j=1} \left( \sin \theta_j \right)^{2n-2}
\end{equation}
for the position probability density, and
\begin{equation}\nonumber
\gamma_{c.s.}(\vec{p})= \frac{2^{2n-2} (2n+D-3)^D \Gamma \left( n+\frac{D-1}{2}\right)}{Z^D \pi^{\frac{D+1}{2}} \Gamma(n)} \frac{(\eta p/Z)^{2n-2}}{(1+\frac{\eta^2 p^2}{Z^2})^{2n+D-1}} \prod^{D-2}_{j=1} \left( \sin \theta_j \right)^{2n-2}
\end{equation}
for the momentum probability density of a D-dimensional hydrogenic circular state with the principal quantum number $n$. Moreover, we have found the values
\begin{equation}\label{DposCS}
\left\langle \rho_{c.s.} \right\rangle = \frac{Z^D \Gamma \left(n-\frac{1}{2} \right) \Gamma \left(2 n + \frac{D-3}{2} \right)}{2^{2n-2} \pi^{\frac{D}{2}} (2n+D-3)^D  \Gamma \left(n \right) \Gamma^2 \left(n+\frac{D-1}{2} \right)}
\end{equation}
and
\begin{equation}\label{DmomCS}
\left\langle \gamma_{c.s.} \right\rangle = \frac{2^{4n+D-4} (2n+D-3)^D \Gamma^2 \left( n+\frac{D-1}{2} \right) \Gamma \left(2n-1 \right) \Gamma \left( 2n+\frac{3D}{2} \right)}{Z^D \pi^{\frac{D+2}{2}} \Gamma^2 \left(n \right) \Gamma \left( 4n+2D-2 \right)}
\end{equation}
for the position and momentum averaging densities of our system. On the other hand, we have also been able to express the position and momentum entropies as
\begin{align}\label{SposCS}
S \left[\rho_{c.s.} \right]&= 2n+D-2 -(n-1) \left[ \psi (n) + \psi \left(n+\frac{D-1}{2} \right)  \right] -D \ln 2\\\nonumber
&\qquad+\ln \left[(2n+D-3)^D \pi^{\frac{D-1}{2}} \Gamma(n) \Gamma  \left(n+\frac{D-1}{2} \right)\right]- D \ln Z
\end{align}
and
\begin{equation}\label{SmomCS}
S \left[\gamma_{c.s.} \right]= A(n,D)+\ln \left[ \frac{2^{D+1} Z^D \pi^{\frac{D+1}{2}} \Gamma(n) }{(2n+D-3)^D \Gamma  \left(n+\frac{D-1}{2} \right) }\right],
\end{equation}
where the constant $A(n,D)$ is given by
\begin{align}\nonumber
A(n,D)=&\frac{2n+D-1}{2n+D-3}-\frac{D+1}{2n+D-2}-(n-1) \psi (n) \\\label{A2}
&\quad -\left(\frac{D+1}{2}\right) \psi \left(n+\frac{D-2}{2} \right) +\left(n+\frac{D-1}{2}\right) \psi \left(n+\frac{D-3}{2} \right)
\end{align}

Finally, from Eqs. (\ref{DposCS})-(\ref{SmomCS}) or from Eqs. (\ref{CposFinal}) and (\ref{CmomFinal}) we have the values
\begin{align}\nonumber
C \left[ \rho_{c.s.} \right] & = \frac{\Gamma  \left(n-\frac{1}{2} \right) \Gamma \left(2n+\frac{D-3}{2} \right)}{2^{2n+D-2} \pi^{1/2} \Gamma \left(n + \frac{D-1}{2} \right)}\\ \label{CposCS}
& \quad \times \exp \left\lbrace 2n+D-2-(n-1) \left[ \psi(n) + \psi \left(n+ \frac{D-1}{2} \right) \right] \right\rbrace
\end{align}
and
\begin{equation}\label{CmomCS}
C \left[ \gamma_{c.s.} \right]= \frac{2^{4n+2D-3}\Gamma \left(n+ \frac{D-1}{2} \right) \Gamma(2n-1) \Gamma \left(2 n + \frac{3D}{2} \right)}{\pi^{1/2} \Gamma(n) \Gamma(4n+2D-2)} \exp \left[ A(n,D) \right]
\end{equation}
for the position and momentum shape complexity of a D-dimensional hydrogenic system in an arbitrary circular state. It is worthwhile remarking for checking purposes that Eqs. (\ref{CposCS}) and (\ref{CmomCS}) reduce to Eqs. (\ref{CposGS}) and (\ref{CmomGS}) in case that $n=1$, respectively, as expected; in this sense we have to use the two following properties of the digamma function: $\psi \left(  2 z \right)= \frac{1}{2} \left[ \psi \left(z \right)+  \psi \left(z +\frac{1}{2}\right) \right]+ \ln 2 $ and $\psi \left(  z+1 \right)=  \psi \left(z \right)+ \frac{1}{z}$.

\section{Numerical study and physical discussion}

Here we discuss the general complexity expressions obtained in the previous Section in terms of (a) the dimensionality for a given circular state (i.e., for fixed $n$), and (b) the principal quantum number $n$ for a given dimensionality.

Let us begin with the dimensional analysis of the position and momentum complexities, $C_D \left[ \rho_{c.s.} \right] $ and $C_D \left[ \gamma_{c.s.} \right]$, given by Eqs. (\ref{CposCS}) and (\ref{CmomCS}), respectively. The resulting position complexity as a function of the dimensionality is drawn at Figure 1 for the ground state (n=1) and the circular states with $n=2$ and $3$. It shows a parabolic growth for all states when $D$ is increasing, being always greater than unity; the minimum value of $C \left[ \rho \right]$ is $\left(\frac{e}{2} \right)^2 =1.847$, what occurs for $D = 2$.

\begin{figure}
\includegraphics[scale=0.4]{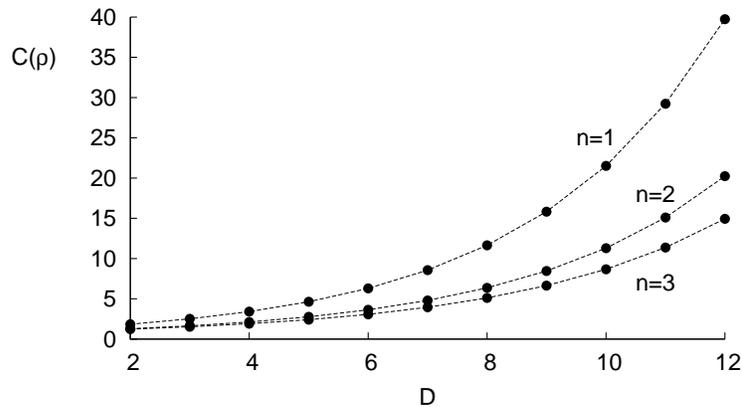} 
\caption{Variation of the shape complexity in position space with the dimension $D$ for three circular states. Atomic units are used.}
\end{figure}

The shape of the momentum complexity (whose minimum value $\frac{2}{5} e^{\frac{3}{2}}=1.793$  corresponds to the case $n=1$ and $D=2$) appears to have a strong ressemblance with the position one, mainly because the two ingredients of each complexity have opposite behaviours when $D$ varies. This is shown in Figure 2, where the Shannon entropies $S \left[ \rho \right]$ and $S\left[\gamma\right]$ as well as the logarithmic values of the position and momentum values of the disequilibrium are plotted for the ground state in terms of $D$. Keep in mind that $C\left[\rho\right] = \exp\left( S \left[ \rho \right]+\ln \left\langle \rho \right\rangle \right)$ in position space and a similar form in momentum space. We observe that the Shannon entropies and the disequilibrium logarithmic measures have opposite behaviours in the two reciprocal spaces, so that the combined exponential effect which gives rise to the corresponding complexities is very similar qualitatively and almost quantitatively. Moreover, it happens that, for a given dimensionality, the relative contribution of the disequilibrium (entropic power) is smaller than that of the entropic power (disequilibrium) in position (momentum) space. This indicates that the relative contribution of the bulk extent of the position (momentum) probability density is more (less) powerful than its average height.

\begin{figure}
\includegraphics[scale=0.4]{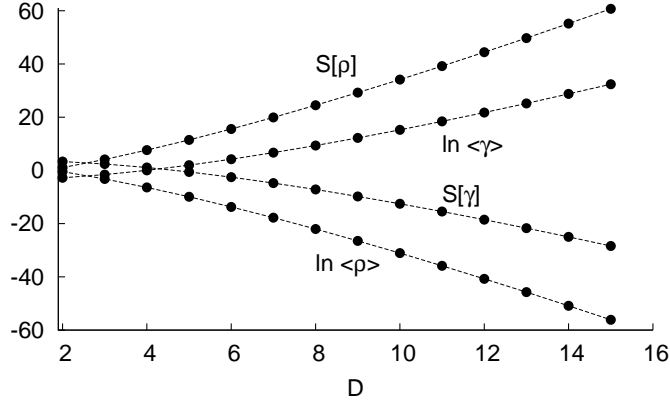} 
\caption{Ground state Shannon entropy ($S(\rho)$, $S(\gamma)$) and disequilibrium ($\left< \rho \right>$, $\left< \gamma \right>$) in position and momentum spaces as a function of the dimension $D$. Atomic units are used.}
\end{figure}

In addition, from Figure 1, we observe that the inequalities
\begin{equation}\nonumber
C_D\left[ \rho_{c.s.}; n=3 \right] < C_D\left[ \rho_{c.s.}; n=2 \right] < C_D\left[ \rho_{g.s.}\right]
\end{equation}
are fulfilled in position space, and similarly in momentum space. This decreasing phenomenon of the complexity for the circular states when the quantum number n is increasing, can be more clearly observed in the left graph of Figure 3 where the values of position complexity for the states with $n=1$-$15$ are given at the dimensionalities $D = 2$, $5$ and $15$. Therein we remark that when the quantum number n is increasing, the radial density behaves so that its maximum height decreases and its spreading increases at different rates in such a way that it overall occurs the phenomenon pointed out by this chain of inequalities; namely, the larger n is, the smaller is the shape complexity of the corresponding circular state.

\begin{figure}
\includegraphics[scale=0.3,angle=270]{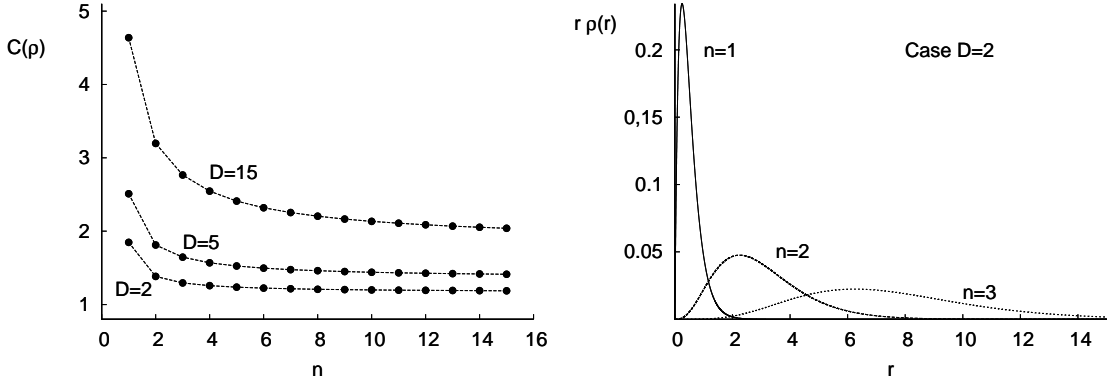}
\caption{Variation of the position shape complexity of circular states with the principal quantum number $n$ for various dimensionalities. (Right)  Radial probability density in position space for various two-dimensional circular states.}
\end{figure}

These dimensional and energetic (quantum number $n$) behaviours of the position complexity turn out to be a delicate overall balance of the average height and the bulk spreading of the system given by its two information-theoretic ingredients: the disequilibrium $\left\langle \rho \right\rangle$ and the Shannon entropic power, respectively.

Now we would like to find the dimensional (i.e., when $D \rightarrow \infty$) limit, and the high energy or Rydberg ($n \rightarrow \infty$) limit of the position and momentum complexities of our system. The former one plays a relevant role in the dimensional scaling methods in atomic and molecular physics \cite{H93}, and the latter one for the Rydberg states which lie down at the region where the transition classical-quantum takes place. The large $D$ limit is closed to (but not the same) the conventional classical limit obtained by $\hbar \rightarrow 0$ for a fixed dimension \cite{H93}.

For the ground state, whose energy is $E_{g.s.}= -2 \left(\frac{Z}{D-1} \right)^2$, the position complexity is, according to Eq. (\ref{CposGS}), $C_D \left[ \rho_{g.s.} \right]= \left(\frac{e}{2} \right)^D$. So that at the pseudoclassical limit, in which the electron is located at a fixed radial distance, the energy vanishes while the position complexity diverges. In momentum space, the shape complexity given by Eq. (\ref{CmomGS}) has the following behaviour
\begin{equation}\label{CRhogsDInf}
C \left[ \gamma_{g.s.} \right] \sim \frac{3^{\frac{3}{2} (D-1)}}{2^{2D-\frac{3}{2}} \sqrt{e}}, \hspace{2cm} D\rightarrow \infty
\end{equation}
for the pseudoclassical limit.

A similar asymptotic analysis of Eq. (\ref{CposCS}) has allowed us to find the following values for the position shape complexity of a general circular state (characterized by the quantum number $n$)
\begin{equation}\label{CRhocsDInf}
C \left[ \rho_{c.s.} \right] \sim \left(\frac{e}{2} \right)^{D+2n-2} e^{(1-n) \psi(n)} \frac{ \Gamma \left(n-\frac{1}{2} \right)}{\sqrt{\pi}}, \hspace{2cm} D\rightarrow \infty
\end{equation}
at the dimensional limit, and the value
\begin{equation}\label{CRhocsNInf}
C \left[ \rho_{c.s.} \right] \sim \left(\frac{e}{2} \right)^{\frac{D-1}{2}}, \hspace{2cm} n \gg 1
\end{equation}
for the circular Rydberg states of a D-dimensional hydrogenic system.

Operating with Eq. (\ref{CmomCS}) in a parallel way, we have obtained the values
\begin{equation}\label{CGamcsDInf}
C \left[ \gamma_{c.s.} \right] \sim \left( \frac{3^{3/2}}{4} \right)^D \frac{3^{2n-\frac{1}{2}} \Gamma(2n-1)}{2^{4n-\frac{5}{2}} \Gamma(n)} e^{(1-n) \psi(n)-\frac{1}{2}}, \hspace{2cm} D\rightarrow \infty
\end{equation}
for the momentum shape complexity of a circular state with quantum number $n$ at the pseudoclassical limit, and the value
\begin{equation}\label{CGamcsNInf}
C \left[ \gamma_{c.s.} \right] \sim \left(\frac{e}{2} \right)^{\frac{D-1}{2}}, \hspace{2cm} n \gg 1
\end{equation}
for the momentum shape complexity of a circular Rydberg state.

Let us also make some comments about the uncertainty products of the position and momentum shape complexities $C\left[ \rho \right]C\left[ \gamma \right]$ for the ground and circular states. The general expressions are readily obtained from Eqs. (\ref{CposGS}) and (\ref{CposCS}) in position space, and from Eqs. (\ref{CmomGS}) and (\ref{CmomCS}) in momentum space. Moreover, this uncertainty product behaves as
\begin{equation}\nonumber
C\left[ \rho_{c.s.} \right]C\left[ \gamma_{c.s.} \right] \sim \left( \frac{e}{2} \right)^{D-1}, \hspace{2cm} n \gg 1
\end{equation}
at the Rydberg limit, and as
\begin{equation}\nonumber
C\left[ \rho_{c.s.}\right]C\left[ \gamma_{c.s.} \right] \sim \left( \frac{3^{3/2} e }{2^3} \right)^{D} \frac{3^{2n-\frac{1}{2}}}{2^{4n-\frac{5}{2}}} \frac{\Gamma^2 \left(n-\frac{1}{2} \right)}{\pi} e^{2n-\frac{5}{2}-2(n-1) \psi(n)}, \hspace{2cm} D \rightarrow \infty
\end{equation}
at the dimensional limit for circular states, where Eqs. (\ref{CRhocsNInf}) and (\ref{CGamcsNInf}), and (\ref{CRhocsDInf}) and(\ref{CGamcsDInf}) have been taken into account. The last expression yields
\begin{equation}\nonumber
C\left[ \rho_{g.s.}\right]C\left[ \gamma_{g.s.} \right] \sim \left( \frac{3^{3/2} e }{2^3} \right)^{D} \left(\frac{3}{2 e^{1/3}} \right)^{\frac{3}{2}}, \hspace{2cm} D \rightarrow \infty
\end{equation}
for the ground state uncertainty product. Finally, for completeness, let us remark that the complexity uncertainty product is always not less than $\frac{e}{2}=1.359$.

\section{Conclusions}

The shape complexity of the hydrogenic system in D-dimensional position and momentum spaces is investigated. This quantity has two information-theoretic ingredients: the disequilibrium and the Shannon entropic power. We have seen that the explicit computation of this complexity is a formidable open task, mainly because the analytical evaluation of the entropic functionals of the Laguerre and Gegenbauer polynomials, $E_1 \left[{\tilde{L}}_{k}^{\alpha} \right]$ and $E_2 \left[{\tilde{C}}_{k}^{\alpha} \right]$, involved in the calculation of the Shannon entropy, has not yet been accomplished.

The general methodology presented here is used to find explicit expressions for the position and momentum complexities of the ground and circular states in terms of the dimensionality and the principal quantum number. Then, these information-theoretic quantities are numerically discussed for various states and dimensionalities as well as in the dimensional and high-lying energy (Rydberg) limits. Briefly, we find that both position and momentum complexities increase (decrease) when the dimensionality (the quantum number of the state) is increasing. This phenomenon is the result of a delicate combined balance of the average height and the bulk spreading of the system given by their two information-theoretic ingredients, the disequilibrium and the entropic power, respectively. Finally, the uncertainty product of the position and momentum shape complexities is examined.

\section*{Acknowledgments}
The authors gratefully acknowledge the Spanish MICINN grant FIS2008-02380 and the grants FQM-481, 1735 and 2445 of the Junta de Andaluc\'ia. They belong to the Andalusian research group FQM-207. S.L.R. and D.M acknwoledge the corresponding FPU and FPI scholarships of the Spanish Ministerio de Ciencia e Innovaci\'{o}n, respectively.

\bibliographystyle{ppcf}
\bibliography{general}

\end{document}